# Adaptive Selection of Codebook Using Assistance Information and Artificial Intelligence for 6G Systems

Denis Esiunin
*Samsung Research*
d.esiunin@samsung.com

Alexei Davydov
*Samsung Research*
av.davydov@samsung.com

*Abstract*—This paper addresses the problem of adaptive codebook (CB) selection for downlink (DL) precoder quantization in channel state information (CSI) reporting. The accuracy of precoder quantization depends on propagation conditions, requiring independent parameter adaptation for each user equipment (UE). To enable optimal CB selection, this paper proposes UE-assisted CB selection at the base station (BS) using reported by the UE statistical channel properties across time, frequency, and spatial domains. The reported assistance information serves as input to a neural network (NN), which predicts the quantization accuracy of various CB types for each served user. The predicted accuracy is then used to select the optimal CB while considering the associated CSI reporting overhead and precoding performance. System-level simulations demonstrate that the proposed approach reduces total CSI overhead while maintaining the target system throughput performance.

## I. INTRODUCTION

Massive Multiple-Input Multiple-Output (M-MIMO) antenna systems have become an integral part of commercial 5-th Generation New Radio (5G NR) deployments in the C-band [1]. In practice, this technology has demonstrated significant advantages, including higher spatial multiplexing gains and improved transmission directivity through the use of advanced beamforming schemes. M-MIMO, along with its evolution for upper mid-bands, is also considered a key solution for meeting the growing capacity demands of future 6-th Generation (6G) cellular networks [2].

The performance benefits of M-MIMO systems are achieved through advanced precoding schemes implemented at the base station (BS) transmitter. Two types of channel state information (CSI) are commonly considered to assist the beamforming operations. The first type relies on channel reciprocity and sounding reference signals (SRS) transmitted by the user equipment (UE). However, its efficiency is limited by uplink (UL) transmission power constraints, making it more suitable for cell-center users. The second, more widely applicable approach is codebook (CB)-based downlink (DL) precoding. In this method, the UE measures the channel response from the BS digital antenna ports using channel state reference signals (CSI-RS). The measured channel is then quantized using a predefined CB and reported to the BS via the UL control channel, enabling DL beamforming.

To support CSI compression, 5G NR specifies various types of CBs that quantize CSI across the spatial, frequency, and Doppler/time domains [3]. The level of compression and resulting UL control overhead are primarily controlled by a set of parameters defining the number of Discrete Fourier Transform (DFT) basis vectors used to represent the quantized CSI. In addition to CB-based methods, 6G system is expected to adopt AI/ML-based approaches for CSI compression, particularly using auto-encoders. This technique offers higher CSI compression levels while maintaining the similar or lower CSI quantization error. Furthermore, different levels of compression can be supported by adapting the number of auto-encoder outputs [4].

A key practical challenge in supporting various CSI quantization is the proper selection of CSI compression level and configuration of the CB parameters to ensure target CSI quantization error and UL control channel overhead. From a network perspective, the quality of CSI quantization is typically unknown at the BS. As a result, practical BS configurations should rely on a common set of CB parameters for all UEs, which may not always provide optimal performance under diverse channel conditions.

The problem of UE-specific CB selection has been discussed in several papers. More specifically, [5] proposes a basic approach based on Line-of-Sight (LoS) and Non-Line-of-Sight (NLoS) channel classification serving as foundation for coarse switching between basic CB types. In [6], a federated reservoir computing framework (CA-FedRC) for 5G NR CB adaptation is introduced, balancing performance and feedback overhead based on some CSI indicators. Using simple link-level channel models, it was shown that dynamic switching of CB under diverse communication conditions can significantly improve throughput performance while reducing UL control channel overhead. However, the corresponding method requires large number of input parameters and some convergence time to make optimal CB selection decision possibly resulting into performance degradation during transient time. In [7] an UE-assistance information-based approach is proposed, leveraging time-variability characteristics of the channel between BS and UE for system parameter selection including basic switching of CBs. The proposed UE assistance information, which corresponds to the channel correlation in time domain, enables 5G network to UE-specifically select feedback parameters (i.e., type of CSI feedback) to maximize both user and system performance. However, the approach proposed in [8] relies solely on traditional single threshold-based CB selection, making it challenging to determine accurate threshold for optimal adaptation. Moreover, time-domain channel properties (TDCP) alone may not be sufficient to determine the optimal CB configurations of all parameters. A more accurate approach should consider other channel characteristics (e.g., in spatial and frequency domains) and employ joint selection method for CB adaptation. In particular, it is well known that CSI compression based on Type-1 CB in 5G NR may experience severe performance degradation in rich scattering environments (i.e., channels with high delay and angular spreads) [8]. However, in high-mobility scenario with lower channel correlation in time-domain, Type-1 CB provides more robust performance and may outperform other CSI quantization methods depending on Doppler spread. This necessitates joint consideration of all channel properties to decide on the optimal CB.

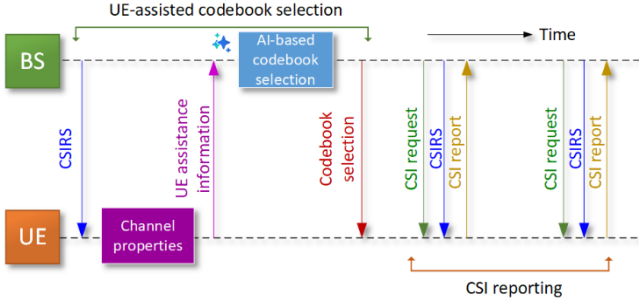

Fig. 1 The considered system, where BS selects CB for the UE using UE-assistance information obtained using reference signals transmitted from the BS antenna ports. The selected CB is then used for CSI reporting

Taking these considerations into account, this paper proposes a more comprehensive UE assistance feedback mechanism that provides channel properties across spatial, frequency, and time dimensions. The assistance information is utilized as input to a neural network (NN) deployed at the BS to predict the CSI quantization accuracy at the UE for all supported CB types. The predicted quantization accuracy also considers channel aging issues associated with CSI measurement and reporting delays (see Fig. 1). Two CB selection strategies are then considered at the BS based on the predicted quantization errors. In the first approach, the CB with the lowest CSI overhead meeting the minimum CSI quantization accuracy requirement is selected for the UE. In the second approach, CB selection takes into account the tradeoff between improved DL precoding, achieved through more accurate CSI, and the UL overhead required for the corresponding CSI transmission.

## II. GENERALIZED PRECODING SCHEME

We consider a general M-MIMO antenna system serving a set of UEs within the coverage area of a BS equipped with cross-polarized and two-dimensional $N_1 \times N_2$ antenna arrays with $P = 2N_1N_2$ digital antenna ports. Each UE periodically or on demand provides the BS with assistance information describing statistical channel properties across all dimensions to enable optimal CB selection. The selected CB based on certain criteria is then configured for the UE to facilitate DL precoding quantization.

For 6G system, we define a generalized precoding matrix $W$ of dimension $P \times (N_f N_t)$, representing beamforming over $N_t$ time slot groups (containing $N_S$ slots) and $N_f$ frequency sub-bands (containing $N_{RB}$ resource blocks) which can be expressed as:

$$W = [\ldots \quad w(k,l) \quad \ldots] = W_1 W_2 (W_f \otimes W_d)^H, \quad (1)$$

where $w(k,l)$ is actual $P \times 1$ precoding vector for $k$-th frequency subband and $l$-th slot group, $(\cdot)^H$ and $\otimes$ represent complex conjugate transpose and Kronecker product respectively. In (1) $W_1$, $W_f$ and $W_d$ correspond to the spatial domain basis, frequency domain basis, and time domain basis respectively. More specifically, $W_1$ is a $P \times 2L$ block diagonal matrix consisting of $2L$ DFT spatial basis vectors obtained from oversampled two-dimensional DFT matrix, while $W_f$ is an $N_f \times M$ matrix composed of $M$ DFT frequency basis vectors obtained from oversampled frequency domain DFT matrix. Similarly, $W_d$ is an $N_t \times T$ matrix consisting of $T$ DFT time basis vectors. The matrix $W_2$ represents the combined space-frequency-time complex coefficients with a dimension $2L \times MT$. Due to sparse structure of the channel after DFT transform, only the $K$ strongest coefficients may be selected for the report from all $2L \times MT$ coefficients in $W_2$. The selected from $W_2$ linear combination (LC) coefficients, along with the corresponding indexes of the chosen DFT basis vectors are reported by the UE.

The precoding structure described in (1) aligns with the Release 18 5G NR CB and is expected to be adopted in 6G as a generalized DL precoding quantization method, enabling flexible quantization adaptation across various dimensions. More specifically, the precoding matrix in (1) can be adjusted for CSI compression over selected dimensions, depending on the scenario. For instance, it can be reduced to spatial-frequency CSI compression (e.g., eType-2 CB in 5G NR) or to spatial-only compression (e.g., Type-2 CB in 5G NR). If CSI compression is performed solely in the spatial dimension using a single DFT vector (i.e., $L = 1$), the DL precoding method becomes aligned with the Type-1 CB in 5G NR systems.

In the generalized DL precoding matrix (1), quantization accuracy can be also adjusted by varying the number of used DFT basis vectors in each compression dimension, i.e., $L$, $M$ and $T$. In particular, increasing the number of DFT basis vectors typically enhances the accuracy of DL precoder quantization. However, this also leads to a higher CSI overhead due to the increased number of coefficients that must be reported in the $W_2$ matrix of dimension $2L \times MT$. Therefore, enabling CB adaptation procedure in 6G system is crucial for achieving optimal precoding performance while avoiding unnecessary UL overhead.

## III. UE-ASSISTANCE INFORMATION

Channel autocorrelation is a statistical metric that describes how rapidly the channel varies over a given dimension, such as space, frequency, or time. It serves as a useful measure for determining the number of required DFT vectors to represent the corresponding DL precoding.

Specifically, channels with high correlation (e.g., LoS channel) typically require fewer DFT vectors to quantize CSI over the corresponding dimension, whereas channels with low correlation require more DFT vectors. For realistic channel models, autocorrelation depends on the propagation conditions of the UE and is difficult to fully characterize using a single parameter [7]. As a result, a general correlation function should be considered for CB adaptation process.

In Release 18 5G NR, the TDCP is introduced to describe channel correlation in the time domain [9]. TDCP is configured using the CSI reporting framework, where each reporting configuration specifies the number of delay values for which normalized autocorrelation (either complex or real-valued amplitudes) is reported. Additionally, the CSI-RS for tracking is used for TDCP measurements are configured as part of the reporting setup. The TDCP metric proposed in [7] is introduced for selecting basic system parameters including type of CB. However, TDCP alone is insufficient for selecting all parameters of generalized precoder. In this paper, we propose extending UE assistance information to additionally include spatial and frequency correlation, enabling more informed selection of CB parameters for generalized precoder in (1), specifically the number of required DFT vectors across the corresponding dimensions. The proposed correlation metrics can be jointly used to predict the accuracy of CSI quantization of different CBs taking into account CSI measurement and reporting delays.

To achieve this objective, the spatial-domain channel autocorrelation with port offsets $\Delta p_1 \in \{0,1,\dots,N_1-1\}$ and $\Delta p_2 \in \{0,1,\dots,N_2-1\}$ in the first and second dimensions, respectively, can be defined as follows:

$$c_p(\Delta p_1, \Delta p_2) = \sum_{n_1,n_2} \sum_f h_{n_1+\Delta p_1, n_2+\Delta p_2}(f,t) h^*_{n_1,n_2}(f,t), \quad (2)$$

where $h_{n_1,n_2}(f,t)$ corresponds to the channel at the $(n_1, n_2)$ antenna port index of the BS, the variable $f$ represents the resource block (RB) index and $t$ is the slot index. Spatial domain channel properties (SDCP), denoted as $\bar{c}_p(\Delta p_1, \Delta p_2)$, is then calculated as normalized correlation function to correlation with zero antenna port offsets, i.e., $\Delta p_1 = \Delta p_2 = = 0$:

$$\bar{c}_p(\Delta p_1, \Delta p_2) = \frac{c_p(\Delta p_1, \Delta p_2)}{c_p(0,0)}. \quad (3)$$

A similar correlation function can be defined in the frequency domain for different RB block offsets denoted as $\Delta f \in \{0,1,\dots,F-1\}$

$$c_f(\Delta f) = \sum_{n_1,n_2} \sum_f h_{n_1,n_2}(f+\Delta f, t) h^*_{n_1,n_2}(f,t). \quad (4)$$

In this case frequency domain channel properties (FDCP), denoted as $\bar{c}_f(\Delta t)$, is then calculated as normalized frequency correlation function to frequency correlation with zero RB offset, i.e., $\Delta f = 0$:

$$\bar{c}_f(\Delta t) = \frac{c_f(\Delta f)}{c_f(0)}. \quad (5)$$

Finally, time domain correlation is calculated for set of time delays $\Delta t \in \{0,1,\dots,Q-1\}$ defined according to [9]

$$c_t(\Delta t) = \sum_{n_1,n_2} \sum_f h_{n_1,n_2}(f, t+\Delta t) h^*_{n_1,n_2}(f,t). \quad (6)$$

TDCP in this case can be defined following current 5G NR specification as normalized time domain correlation to correlation with zero delay, i.e., $\Delta t = 0$:

$$\bar{c}_t(\Delta t) = \frac{c_t(\Delta t)}{c_t(0)}. \quad (7)$$

For calculation of SDCP and FDCP, the conventional CSI-RS can be reused without creating extra reference signals and reported to the BS using the same approach as TDCP.

## IV. AI-BASED QUANTIZATION ERROR PREDICTION

### A. Aging-Aware Generalized Cosine Similarity

In this paper, it is assumed that UE provides assistance information to the BS in the form of $\bar{c}_p(\Delta p_1, \Delta p_2)$, $\bar{c}_f(\Delta f)$ and $\bar{c}_t(\Delta t)$ reports to support prediction of the DL precoding quantization accuracy for different CBs. To this end, Generalized Cosine Similarity (GCS) is widely used to characterize CSI quantization errors for both CB and Artificial Intelligence (AI) driven compression methods [10]. GCS serves as an intermediate indicator for evaluating CSI compression and recovery accuracy, where a higher GCS value typically reflects improved performance. Consequently, this paper proposes GCS as a metric for predicting the accuracy of DL precoding across various CBs.

To account DL precoding aging in the presence of UE mobility, the conventional GCS metric should be extended to incorporate CSI feedback measurement and reporting delays. We propose aging-aware GCS (AGCS), which extends the standard GCS formulation to capture the impact of channel variation over time. Specifically, for a given CB $\mathbb{C}$ defined by the parameters set $L$, $M$ and $T$ discussed in Section II, AGCS denoted as $\rho_\delta(\mathbb{C})$ is defined as follows:

$$E\left\{\frac{1}{N}\sum_{k=1}^{N_f}\sum_{l=1}^{N_t}\sum_{m=1}^{N_{RB}}\sum_{n=1}^{N_s} \frac{|v(k,l,m,n)^H w_\delta(k,l)|}{||v(k,l,m,n)|| \cdot ||w_\delta(k,l)||}\right\}. \quad (8)$$

In this formulation:

- $||\cdot||$ corresponds to the $\ell_2$-norm,
- $E\{\cdot\}$ represents the expectation over different channel realizations of the UE,
- $k, l, m$ and $n$ define frequency subband index, slot group index, RB index in the subband and slot index in the slot group respectively,
- $v(k,l,m,n)$ is the ideal DL precoder,
- $w_\delta(k,l)$ is the quantized DL precoder obtained according to expression (1) taking into account possible CSI measurement and reporting delay $\delta$,
- $N_{RB}$ and $N_s$ are number of RBs in the frequency sub-band and number of slots in the slot group,

- $N = N_f N_t N_{RB} N_s$ denote the total number of samples to compute AGCS.

Unlike the conventional GCS, the ideal and quantized DL precoder in equation (8) are evaluated with consideration of granularities of the DL precoder updates not only in the frequency, but also in the time domain. Additionally, the parameter $\delta$ effectively accounts for CSI measurement and reporting delays, capturing channel aging effects caused by Doppler. This is particularly important for CSI quantization schemes that do not perform compression in Doppler/time domain for CSI prediction.

*B. AGCS Prediction Problem*

In this paper, we address the AGCS prediction problem for CB $\mathbb{C}$, taking into account CB parameters $L, M, T$ and the impact of CSI measurement and reporting delay $\delta$.

Our objective is to achieve an accurate estimation of the DL precoding quantization errors, reflected by $\rho_\delta(\mathbb{C})$, based on the known channel properties $\bar{c}_p, \bar{c}_f, \bar{c}_t$ reported to the BS by the UE. The predicted value $\hat{\rho}_\delta(\mathbb{C})$ is subsequently employed for optimal CB selection at the BS, following the schemes outlined in Section V.

The desired AGCS prediction function $\mathcal{F}$ is parametrized by set of parameters $\Theta$ and can be expressed as follows:

$$\hat{\rho}_\delta(\mathbb{C}) = \mathcal{F}(\mathbb{C}, \bar{c}_p, \bar{c}_f, \bar{c}_t; \Theta). \quad (9)$$

Consequently, the problem lies in finding the parameter set $\Theta$ that provides accurate estimation of AGCS using available UE assistance information. Since the expression in (9) fundamentally involves learning a complex function $\mathcal{F}$, we adopt in this paper NNs as the AGCS estimator.

*C. Training Loss Function for NN*

Since the goal of NN-based function in (9) is to predict the AGCS, we formulate the learning task as a regression problem using a supervised learning approach. To train the NN model, we use mean squared error (MSE) as the loss function. For each predicted AGCS sample, the loss function is defined as follows:

$$\mathcal{L}\left(\rho_\delta(\mathbb{C}), \hat{\rho}_\delta(\mathbb{C})\right) = \left\|\rho_\delta(\mathbb{C}) - \hat{\rho}_\delta(\mathbb{C})\right\|^2$$
$$= \left\|\rho_\delta(\mathbb{C}) - \mathcal{F}(\mathbb{C}, \bar{c}_p, \bar{c}_f, \bar{c}_t; \Theta)\right\|^2. \quad (10)$$

The objective of learning process is to find a set of NN model parameters that minimizes the average loss over the entire dataset, which can be expressed as follows:

$$\min_{\Theta} \frac{1}{D} \sum_{d=1}^{D} \mathcal{L}(\rho_\delta(\mathbb{C}), \hat{\rho}_\delta(\mathbb{C})), \quad (11)$$

where $D$ is the size of data set, the details of which will be provided in the Section VI and $\Theta$ are the trained model parameters.

*D. Network Architecture*

For AGCS prediction, we adopt a simple NN-based on a fully connected (FC) architecture composed of $U$ layers. We denote the input layer as 0-th layer and the output layer as $(U+1)$-th layer. Each hidden layer contains $(N_1 N_2 + F + Q)/2$ neurons. The output layer contains $G$ nodes, representing the $G$ possible CBs candidates for which AGCS is evaluated. ReLU activation function is employed in the output layer as well as in the hidden layers. For improved performance and reduced complexity, AGCS prediction can also utilize convolutional NN or other advanced architectures. However, even with the simple FC model considered in the paper, accurate prediction results were achieved.

## V. CODEBOOK SELECTION SCHEMES

In this section, we propose potential strategies for selecting a CB based on the estimated AGCS of each candidate CB. We assume that all candidate CBs $\mathbb{C}$ are ordered according to their CSI reporting overhead in increasing order.

The first approach proposed in the paper involves selecting the CB with the lowest CSI overhead that meets the minimum CSI quantization accuracy requirement, i.e.,

$$\mathbb{C}_{\text{opt}} = \text{argmax}(\hat{\rho}_\delta(\mathbb{C}) \geq \rho_{min}, \text{'first'}). \quad (12)$$

In (12), the AGCS threshold $\rho_{min}$ for CB selection is common across all candidate CBs and can be selected based on the target transmission scheme in use. For instance, in single-user MIMO (SU-MIMO) scheme, $\rho_{min}$ can be set to a lower value than in multi-user MIMO (MU-MIMO) transmission. This adjustment is necessary because MU-MIMO requires better accuracy for intra-cell interference suppression, which demands a higher threshold value. Additionally supporting a larger number of MU-MIMO layers further increases the requirements on quantization accuracy, necessitating a higher threshold $\rho_{min}$.

In the second approach, the reference CB (e.g., Type-1), denoted as $\mathbb{C}_r$, is used by the BS as default option. If the AGCS difference between a candidate CB $\mathbb{C}$ and the reference CB $\mathbb{C}_r$ becomes better than pre-determined threshold, the candidate CB with largest AGCS is selected. This condition is expressed as follows:

$$\mathbb{C}_{\text{opt}} = \text{argmax}(\hat{\rho}_\delta(\mathbb{C})) \text{ subject to}$$
$$\hat{\rho}_\delta(\mathbb{C}) - \hat{\rho}_\delta(\mathbb{C}_r) \geq \rho_0(\mathbb{C}), \quad (13)$$

where $\rho_0(\mathbb{C})$ is CB dependent threshold accounting the difference in CSI overhead among candidate CBs. Compared to the approach in (12), this method introduces a CB-specific threshold $\rho_0(\mathbb{C})$ to better reflect the varying CSI overheads. As a result, this approach effectively balances the trade-off between improved DL precoding, achieved through more accurate CSI, and the UL overhead required for the corresponding CSI transmission.

## VI. SIMULATION RESULTS

*A. Channel Model and Dataset*

In simulations M-MIMO system deployed in upper mid band with central frequency of 7GHz and 100MHz bandwidth is considered. The BS is equipped with 256 cross-polarized digital antenna ports ($\pm 45^0$ slant) arranged into two-dimensional array with dimensions $N_1 = 16$ ports in the horizontal dimension and $N_2 = 8$ in the vertical dimension. Each antenna port is connected to 4 vertical antenna elements with antenna spacing of 0.6 wavelength creating $32 \times 16$ antenna array at the BS. UE has 2 pairs of cross-polarized antennas.

The channels for data set are generated using system-level simulator for 3GPP Urban Macro (UMa) scenario with inter-site distance (ISD) of 500m. Both LoS and NLoS propagation conditions are considered according to 3GPP evaluation methodology. The parameters of the CBs for NN training are provided in Table I, where for simplicity DL quantization schemes are considered without CSI compression in Doppler/time domain.

### B. AGCS Prediction Accuracy Evaluations

We first evaluate the accuracy of AGCS prediction in the scenarios with low UE mobility. Table II illustrates the achieved AGCS prediction accuracy, characterized by MSE, averaged across two MIMO layers. As can be seen the NN is able to accurately estimate DL precoder quantization accuracy based on the proposed UE-assistance information. It is interesting to note that SDCP is sufficient information to predict AGCS for CB with coarse spatial domain quantization using single DFT vector, i.e., $L = 1$. However, for other CB types supporting CSI quantization in both spatial and frequency domains, i.e. $L > 1$ and $M > 1$, FDCP should be additionally used by NN at the BS for improved prediction performance.

Table II also presents the results for AGCS prediction in a mixed-mobility scenario, where the CSI measurement and reporting delay was set to 5ms. The evaluation considered mobility speeds of 3km/h for all indoor UEs (80%) and 30km/h for all outdoor UEs (20%). The results indicate that SDCP and FDCP alone are insufficient for accurately predicting AGCS across all CB types. However, incorporating TDCP information in NN as input parameter significantly improves the AGCS prediction accuracy. Additionally, Fig. 2 shows the detailed distribution of AGCS of the CBs for low-mobility and high-mobility users. It is observed that for most of outdoor UEs with high mobility containing LoS and NLoS cases, DL precoding based on the first CB configuration in Table I with $L = 1$ demonstrates better AGCS performance than other CB types. However, the AGCS performance trend is different for low-mobility users. This observation aligns with the previous findings in [7]

TABLE I. QUNATIZATION PARAMETERS OF CODEBOOKS

| CB Case $\mathbb{C}$ | Quantization Parameters | | | 5G NR CB |
|---|---|---|---|---|
| | L | M | K | |
| 0 | 1 | - | - | Type-1 |
| 1 | 2 | 5 | 10 | eType-2 |
| 2 | 2 | 5 | 20 | eType-2 |
| 3 | 12 | 5 | 60 | eType-2[a] |
| 4 | 12 | 9 | 216 | eType-2[a] |

[a.] New parameters for M-MIMO with 256 ports

TABLE II. AGCS MSE ($\times 10^{-3}$) FOR NN-BASED PREDICTION

| CB Case $\mathbb{C}$ | 0 | 1 | 2 | 3 | 4 |
|---|---|---|---|---|---|
| | Low-Mobility Scenario | | | | |
| SDCP | 3.5 | 6.2 | 7.1 | 16.4 | 20.1 |
| FDCP | 4.4 | 5.2 | 5.6 | 4.2 | 4.0 |
| TDCP | 5.7 | 9.2 | 9.4 | 15.0 | 16.6 |
| SDCP+TDCP | 3.8 | 5.7 | 6.4 | 12.3 | 15.3 |
| FDCP+TDCP | 4.4 | 5.0 | 5.3 | 3.8 | 3.4 |
| SDCP+FDCP | 3.0 | 3.1 | 3.3 | 3.4 | 3.7 |
| | Mixed-Mobility Scenario | | | | |
| SDCP+FDCP | 4.1 | 4.7 | 5.0 | 6.5 | 7.7 |
| SDCP+FDCP+TDCP | 3.8 | 3.8 | 3.9 | 3.3 | 3.2 |

which indicate that the performance of eType-2 DL precoding is more sensitive to Doppler effects.

### C. Feature Importance Analysis

One potential drawback of supporting UE-assistance is the additional CSI overhead. This overhead can be minimized by pruning certain components in the SDCP, FDCP and TDCP reports that have minimum impact on the prediction accuracy. This approach, known as feature importance analysis, is a common technique used in machine learning to reduce dimensionality of NN by identifying the input variables that have the greatest influence on a model's predictions.

In the context of CB adaptation through AGCS prediction, we adopt permutation importance [11] – a powerful and model-agnostic technique that estimates feature importance by measuring the effect of interleaving individual feature values on NN model performance. By randomly permuting feature values and observing the resulting performance degradation, we can assess the significance of each feature – with larger performance drops indicating greater importance. Our analysis, presented in Fig. 2 for low-mobility case, shows in Table III that approximately 20%-40% of the calculated UE-assistance information is typically sufficient to achieve good AGCS prediction accuracy. However, further sub-sampling of UE-assistance information, contained in $\bar{c}_p, \bar{c}_f$ and $\bar{c}_t$, significantly degrades the AGCS prediction performance.

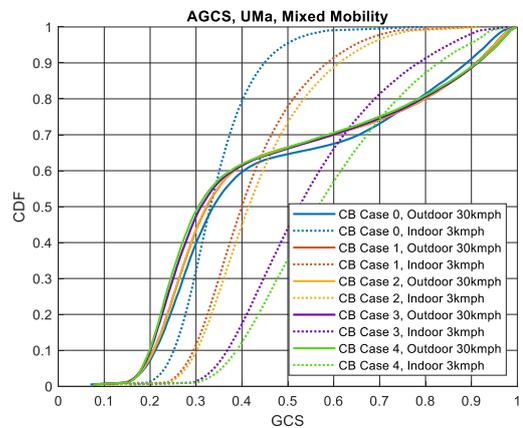

Fig. 2 AGCS distribution for different CBs

TABLE III. MSE (×10$^{-3}$) OF AGCS WITH REDUCED OVERHEAD

| Overhead Reduction in UE-Assistance Info | CB Case ℂ | | | | |
|---|---|---|---|---|---|
| | 0 | 1 | 2 | 3 | 4 |
| 0% | 3.0 | 3.1 | 3.3 | 3.4 | 3.7 |
| 40% | 3.1 | 3.1 | 3.3 | 3.3 | 3.5 |
| 60% | 3.7 | 3.9 | 4.2 | 3.4 | 3.6 |
| 80% | 3.8 | 4.0 | 4.2 | 4.3 | 4.3 |
| 95% | 5.1 | 5.4 | 5.6 | 5.8 | 5.3 |

*D. System-level Evaluations*

A UE-specific CB selection schemes based on the predicted AGCS were also evaluated in the system level simulations (SLS), where the accuracy of the corresponding CB selection is directly reflected in the PDSCH precoding efficiency. For the evaluation, the first and the second CB selection strategies described in Section V were applied to the SU-MIMO scenario.

In the first CB selection strategy, the CB with the lowest CSI overhead that provided the predicted AGCS above the threshold $\rho_{min} = 0.55$ was selected for the UE. If the predicted AGCS for all CB types fell below this threshold, the CB with the highest predicted AGCS was chosen instead. In the second approach the AGCS thresholds for CB selection were set equal to $\rho_0$ = {0.04, 0.045, 0.1, 0.25} to provide the balance between DL precoding performance and CSI overhead.

The user-perceived throughput (UPT) results and CSI overhead comparison are presented in Fig. 3, where CB case 4, offering the minimum quantization loss for DL precoding, was used as baseline. For performance comparison, the ideal LoS/NLoS-based CB adaptation from [5] was also evaluated, assigning LoS UEs to CB case 0 and NLoS UEs to CB case 4. The results demonstrate that both proposed UE-specific CB selection approaches, based on predicted AGCS, significantly outperform fixed CB case 0 DL precoding in terms of average and 5%-tile UPT. They also exceed the performance of the ideal LoS/NLoS based CB adaptation by 5% on average, while limiting the 5%-tile UPT loss to less than 3% with reduced CSI overhead. Moreover, they achieve performance comparable to the baseline DL precoding based on CB case 4 while reducing CSI overhead by approximately 45% - 53%.

Additionally, the results in Fig. 3 indicate that the second CB selection approach achieves better CSI overhead saving. The improvement stems from a more intelligent CB selection across the entire AGCS range including low values. In contrast, the first CB selection strategy does not adaptively regulate AGCS for the values below the threshold $\rho_{min}$. As the result the specific selection of the CB in the corresponding region may not be always optimal from CSI overhead perspective, particularly for low AGCS values.

VII. CONCLUSIONS

This paper presents a novel approach for adaptive DL CB selection in 6G systems. The proposed method leverages assistance information from the UE, which captures channel correlation properties across spatial, frequency, and time dimensions. This information is utilized by a NN at the BS to predict the DL precoder quantization accuracy of various CBs, considering the channel aging effects induced by user mobility. System-level simulations demonstrate that the proposed approach effectively reduces total CSI overhead while maintaining the target system throughput performance.

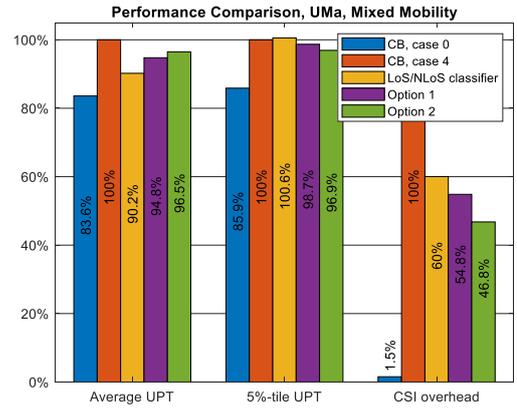

Fig. 3 SLS throughput and CSI overhead results